# BN white graphene with 'colorful' edges—the energies and morphology

Yuanyue Liu, Somnath Bhowmick, and Boris I. Yakobson

Department of Mechanical Engineering & Materials Science, Department of Chemistry,
and the Smalley Institute for Nanoscale Science and Technology,
Rice University, Houston, TX 77005, USA
E-mail: biy@rice.edu

Interfaces play a key role in low dimensional materials like graphene or its boron nitrogen analog, white graphene. The edge energy of h-BN has not been reported as its lower symmetry makes it difficult to separate the opposite B-rich and N-rich zigzag sides. We report unambiguous energy values for arbitrary edges of BN, including the dependence on the elemental chemical potentials of B and N species. A useful manifestation of the additional Gibbs degree of freedom in the binary system, this dependence offers a way to control the morphology of pure BN or its carbon inclusions, and to engineer their electronic and magnetic properties.

Current attention to the atomic mono-layers of carbon—graphene, and especially the need to introduce an electronic gap in this gapless sheet, has brought to the spotlight its sibling 2D-material of hexagonal boron nitride, h-BN [1]. Its bulk phase (often called white graphite) with weak inter-layer bonding permits separation into individual sheets of white graphene. Furthermore, analogous to carbon nanotubes, tubular BN structures have been theoretically predicted [2], consequently synthesized [3, 4], and display a number of properties of fundamental and practical importance [5, 6].

Morphologically alike to the honeycomb graphene and even with quite similar bond length (instead of 2.46 Å in C, lattice parameter in BN is $l$ = 2.51 Å, henceforth used as a unit of length), the chemical alternation of B and N atoms causes the ionic nature of this distinctly different, insulating crystal. This makes BN interesting not only on its own, but especially as a counterpart to carbon graphene. Recent work has advanced the connection between the white and black graphene beyond the sheer analogy towards producing actual layers of hybridized BN and graphene domains [7, 8]. These experiments suggest the intriguing possibility of interfacing the white graphene (BN) with the "black" (C) within the same monolayer plane.

Energy analysis suggests [9] that BNC mixtures should separate into immiscible BN and C. If indeed the distinct phases of BN and C coexist as the phase-separate epitaxial domains, their shape must be guided by the thermodynamic interface preferences. The corresponding quantity is the interface energy $\Gamma(\chi)$ for different orientation angles, $\chi$. This is of course closely related with a more basic property, the energy of pristine BN edges, $\gamma(\chi)$. Here we investigate how these interface energies can be calculated, in spite of low symmetry of the lattice, and reveal their variability with chemical potential of constituents. We further define the equilibrium shape of either free BN-clusters or of the inclusions of C in BN and vice versa, and discuss—rather briefly—what effects does it have on their electronic and magnetic properties.

The BN/C interfaces and the free BN edges can be of different kinds, depending on the edge-cut direction. In analogy to graphene, the basic edges are either along the zigzag (Z) atomic motif or along the armchair (A) pattern [10, 11]; in between, a variety of chiral edges can be cut at some angle $\chi$. At this however the similarity ends, as the distinctly different physics of BN edges is caused by the lack of inversion symmetry in the lattice and its binary composition. We will see that the former complicates the definition of the edge energies, while the latter may offer an advantage: having two chemical constituents instead of one adds extra degree of freedom in the Gibbs phase rule, that is a freedom to balance the chemical potential between B and N, which permits control of the shapes, properties, and can be of practical interest.

In spite of these fundamental differences, we can initially use the energy decomposition ansatz [11], to represent the edge energy as a sum over the atoms of A and Z types. For BN though, an immediate generalization must be made, to account for the fact that its zigzag edges are "chemically unbalanced" and expose either all B or all N atoms, henceforth ZB or ZN. Consequently, the nontrivial chiral angles range doubles to 60°, instead of 30° for graphene or carbon tubes. Choosing the armchair BN-bond direction as reference for $\chi = 0$, we obtain [11],

$$\gamma(\chi) = |\gamma| \cos(\chi + C) \tag{1}$$

where $|\gamma| = 2(\gamma_A^2 + \gamma_{Zx}^2 - \sqrt{3}\gamma_A\gamma_{Zx})^{1/2}$ and $C = \text{sgn}(\chi)\cdot\arctan(\sqrt{3} - 2\gamma_{Zx}/\gamma_A)$, with the subscript x = N at $-30° < \chi < 0$ or x = B at $0 < \chi < 30°$. This equation determines the energy of arbitrary edge, as long as the basic energies $\gamma_A$, $\gamma_{ZB}$, and $\gamma_{ZN}$ are known (that is for $\chi = 0$ and $\pm 30°$). We then proceed finding these important quantities.

A common way to find a surface or edge energy is to compute the total slab energy [12, 13], subtract the energy of equivalent material in its bulk form, and assign the rest to the edges [14]. For BN ribbons with the basic edges of length L (in units of *l*) one has, Fig. 1 a-b,

$$\gamma_A = (E_= - M_{BN}\mu_{BN})/2L, \quad \text{and} \quad \gamma_Z \equiv (\gamma_{ZB} + \gamma_{ZN})/2 = (E_= - M_{BN}\mu_{BN})/2L. \tag{2}$$

Here $E_=$ is the total energy of a ribbon (subscript "=") oriented along either A or Z direction, $M_{BN}$ is a number of constituent BN-pairs, and $\mu_{BN} = \mu_B + \mu_N$ is their energy in the BN sheet. It can be chosen as zero level, while the chemical potentials for individual species can vary as $\mu_{B,N} = \frac{1}{2} \mu_{BN} \pm \mu$. This approach, straightforward in case of graphene, gets however only "half job" done for BN. Although computations do yield $\gamma_A = 1.91$ eV and $\gamma_Z = 3$ eV, the latter is just a nominal average of the B- and N-rich edges (Fig. 1b), while the true physical values remain elusive: only found is the sum of the opposite sides, but they seem to remain inseparable. This general issue with materials of lower symmetries, especially non-invariant upon inversion like BN, has been underscored by Cahn [15]. Having such problem encountered earlier for 3D-semiconductors, we have overcome it by considering the polyhedra with identical faces [13]. For a simpler 2D-case of BN, one can consider the triangles surrounded by either all-ZB or all-ZN, Fig. 1c-d. Subtracting the material "cost" from the total energy of such triangle, and omitting the insignificant contribution from the corners (which does not scale with size L), one recovers their true edge energies. The triangle inversion alters the excess of B to the excess of N around the perimeter,

$$(E_\blacktriangle - M_{BN}\mu_{BN} - L\mu_B)/3L = \gamma_{ZB}(\mu) = \gamma^o_{ZB} - \mu/3, \tag{3a}$$

$$(E_\blacktriangledown - M_{BN}\mu_{BN} - L\mu_N)/3L = \gamma_{ZN}(\mu) = \gamma^o_{ZN} + \mu/3, \tag{3b}$$

For a ▲-triangle of size L, direct counting yields $M_{BN} = \frac{1}{2} L(L+3)$, one extra N omitted as a corner correction, and L extra B atoms. The latter is important as it means that the edge energy depends linearly on the chemical potential, as the right hand side specifies. For the inverted ▼-triangle, B and N are interchanged, Eq. 3b. Direct energy calculations [16] for a series of B-rich and N-rich triangular clusters of increasing size yield the data in Fig. 1e. Clearly linear plots show that the size is sufficient, and their slopes determine the edge energies. By choosing the elemental chemical potentials as equal, $\mu_B = \mu_N = \frac{1}{2}\mu_{BN}$, that is $\mu = 0$, we determine the values $\gamma^o_{ZB} = 3.26$ eV and $\gamma^o_{ZN} = 2.72$ eV. The average $[\gamma_{ZB}(\mu) + \gamma_{ZN}(\mu)]/2 = 2.99$ eV is independent of $\mu$ and agrees with $\gamma_Z$ of Eq. (2) to 0.1%, showing excellent consistency. What one gained is now well defined energies for the basic edges of BN, at arbitrary chemical conditions:

$$\gamma_A = 1.91, \ \gamma_{ZB} = 3.26 - \mu/3, \ \gamma_{ZN} = 2.72 + \mu/3, \text{ all in eV.} \tag{4}$$

Together with Eq. (1), Eq. (4) describes the BN edge of arbitrary direction, including the basic A, ZB, ZN, and the chiral types, at any chosen chemical potential $\mu$. This in turn enables the use of Wulff construction [12] to easily determine the equilibrium shapes of pristine BN clusters, at different conditions. We note that, unlike in case of mono-elemental graphene, [10] the shapes can vary broadly with the chemical potential $\mu$.

More important than the isolated clusters are possible epitaxial BN inclusions into C-graphene, or vice versa, the inclusions of black-graphene C into white-graphene BN matrix, in their hybrid mono-layers [7, 8]. For these systems of emerging interest, we go on now to determine the energies $\Gamma(\chi)$ of BN/C interfaces (Fig. 2a), which control the equilibrium morphology of such hybrids. The above approach can be undertaken again, in principle. In practice however, computations for the inclusion of BN in C (or C in BN) become exceedingly expensive. It is more efficient then to use already determined pristine edge values of Eq. (4) and adjust them by the appropriate binding energies $E_{BN-C}$ at interface,

$$\Gamma_{BN/C} = \gamma_{BN} + \gamma_C - E_{BN-C} \tag{5}$$

For clarity the material-subscripts are included here, while elsewhere the blank $\Gamma$ or $\gamma$ refer to the BN/C interface or BN pristine edge. It is opportune that the binding energy, as just a difference between the joint and detached counter-sides (similar to the work of adhesion or cleavage energy) does not depend on $\mu$, and its calculation is unobstructed by the low symmetries. Then, since $\gamma_{BN}$ depends on chemical potential, so does the $\Gamma_{BN/C}$. We also include as example the hydrogen-passivated edges BN-H, Fig. 2b. The data in Fig. 2c show that the termination of the dangling bonds significantly reduces the energy in both cases, compared to pristine edge, as expected. Still, the chemical balance between B and N (μ value) controls the interface energy in the same way as for pristine edge.

As a conductor-in-insulator hybrid [17], graphene embedded in BN matrix appears more interesting than the opposite. Any practical realization would greatly benefit from some degree of control of domain patterns. This poses an important question of what equilibrium shapes to expect, as determined by Wulff construction, based in turn on the interface energy $\Gamma(\chi)$? Fig. 3 shows the basic interface energies along with the computed Wulff constructions, for a range of chemical conditions, quantified by the value of μ. Modulation of the latter should allow one to broadly vary the shapes of inclusions, in striking contrast to pure graphene where the equilibrium islands are nearly hexagonal or more rounded polygons. Notably, non-intuitive shapes such as sharp triangles become preferred at N-rich (left) or B-rich (right) conditions, Fig. 3. In the middle, the Wulff construct is hexagon, with truncated-rounded polygons at the transient μ values.

Along with the shape, other properties such as electronic gap and especially magnetism also change. The magnetism originates from π-electrons of carbon and is localized at the Z-edge. Computations show [16] that triangle graphene domains are ferromagnetic with the total spin equal to the half of excess number of B or N atoms around the perimeter, obeying the Lieb's theorem [18]. The thin arrows in Fig. 3b-f show the cumulative spin around the borders of different domains: triangles have the largest magnetic moment, reduced as the shapes get truncated, and then vanishing for non-magnetic armchair-edge hexagon. The calculated change of magnetic moment per perimeter unit |**μ**| is summarized in Fig. 3a by a dotted line. While the details of magnetism and electronics of the emerging islands deserve a separate study, beyond the scope of this report, we mention a few basic features. As example, we computed the spin density for the ferromagnetic triangles, either B-rich (Fig. 4a) or N-rich (Fig. 4b) at the borders, with the densities of opposite spin shown in opposite shades of gray. If these mutually-inverted triangles were to coexist in the same lattice, their magnetic moments would be antiparallel.

Due to quantum confinement, the intrinsically semi-metallic graphene isles gain the characteristics of a quantum dots (QD, similar to vacancy clusters in graphane [19]): Fig. 4 c-d show the computed flat, dispersionless bands with significant HOMO-LUMO distances. The band gap $E_g$ scales with the size M of the isle-QD as $E_g \sim 1/\sqrt{M}$, following the trend for confined Dirac fermions[19, 20]. Further, in contrast to nonmagnetic hexagonal QD, a magnetic triangular isle displays distinctly different energy gaps for the spin majority and minority bands. Calculations show that the B-terminated QD have larger band gaps for spin majority than minority, while in the N-terminated QD the band gaps are in the reverse order, smaller for majority-spin bands and greater for minority.

A systematic way discussed above allows one to determine the energies $\gamma(\chi)$ of arbitrary h-BN edges, for the basic directions given in Eq.(4). The dependence on chemical potential difference $\mu$ between B and N species suggests that the equilibrium nanoparticles should be triangular (a counterintuitive shape of larger perimeter) with zigzag edges, except the narrow middle range of chemical conditions where more compact hexagons with armchair edge dominate. In fact, experimental observations [21-23] and particularly the shape evolution between hexagonal and triangular nanoplates [24] strongly support the notion of morphology control discussed above. Making further connection to BN nanotubes, their chiral distribution is not detailed, but believed in recent reports [5, 25] to be mostly zigzag; this contrasts to carbon tubes, and can be explained by the preference of zigzag edge (Fig. 2c) in either boron-rich or nitrogen-rich conditions. For graphene isles in BN, the BN/C interface energies are also unambiguously computed, as $\Gamma_A = 0.56$ and $\Gamma_{ZB} = (0.95 - \mu/3)$, $\Gamma_{ZN} = (0.38 + \mu/3)$, Fig. 3a. Their variability suggests a way to selectively synthesize the inclusions of desirable shapes, with triangular quantum-dots islands displaying largest magnetic moment. Additional work can clear the ways to use the balance between the B, N and C for design and growth of their hybridized domains with desirable physical properties. A concluding disclaimer can hardly be more precise than a quote from Conyers Herring's classics:[12] "Although the interpretation of experiments in such fields as the shapes of small particles and the thermal etching of surfaces usually involves problems of kinetics rather than mere equilibrium considerations, it is suggested that a knowledge of the relative free energies of different shapes or [edge] configurations may provide a useful perspective."

***

[1]     A. Nag, K. Raidongia, K. P. S. S. Hembram, R. Datta, U. V. Waghmare, and C. N. R. Rao, Acs Nano **4**, 1539 (2010).

[2]     A. Rubio, J. L. Corkill, and M. L. Cohen, Physical Review B **49**, 5081 (1994).

[3]     N. G. Chopra, R. J. Luyken, K. Cherrey, V. H. Crespi, M. L. Cohen, S. G. Louie, and A. Zettl, Science **269**, 966 (1995).

[4]     D. Golberg, Y. Bando, M. Eremets, K. Takemura, K. Kurashima, and H. Yusa, Applied Physics Letters **69**, 2045 (1996).

[5]     D. Golberg, Y. Bando, Y. Huang, T. Terao, M. Mitome, C. Tang, and C. Zhi, ACS Nano **4**, 2979 (2010).

[6]     H. Zeng, C. Zhi, Z. Zhang, X. Wei, X. Wang, W. Guo, Y. Bando, and D. Golberg, Nano Letters **10**, 5049 (2010).

[7]     L. Ci, L. Song, C. Jin, D. Jariwala, D. Wu, Y. Li, A. Srivastava, Z. F. Wang, K. Storr, L. Balicas, F. Liu, and P. M. Ajayan, Nature Materials **9**, 430 (2010).

[8]     L. Song, *et al.*, Nano Letters **10**, 3209 (2010).

[9]     K. Yuge, Physical Review B **79**, 144109 (2009).

[10]    C. K. Gan, and D. J. Srolovitz, Physical Review B **81**, 125445 (2010).

[11]    Y. Liu, A. Dobrinsky, and B. I. Yakobson, Physical Review Letters **105**, 235502 (2010).

[12]    C. Herring, Physical Review **82**, 87 (1951).

[13]    K. Rapcewicz, B. Chen, B. Yakobson, and J. Bernholc, Physical Review B **57**, 7281 (1998).

[14]    S. Jun, Physical Review B **78**, 073405 (2008).

[15]    E. Arbel, and J. W. Cahn, Surface Science **51**, 305 (1975).

[16]    See supplemetal material.

[17]    J. M. Pruneda, Physical Review B **81**, 161409 (2010).

[18]  E. H. Lieb, Physical Review Letters **62**, 1201 (1989).

[19]  A. K. Singh, E. S. Penev, and B. I. Yakobson, Acs Nano **4**, 3510 (2010).

[20]  M. V. Berry, and R. J. Mondragon, Proceedings of the Royal Society of London. A. Mathematical and Physical Sciences **412**, 53 (1987).

[21]  C. Jin, F. Lin, K. Suenaga, and S. Iijima, Physical Review Letters **102**, 195505 (2009).

[22]  J. C. Meyer, A. Chuvilin, G. Algara-Siller, J. Biskupek, and U. Kaiser, Nano Letters **9**, 2683 (2009).

[23]  L. Q. Xu, J. H. Zhan, J. Q. Hu, Y. Bando, X. L. Yuan, T. Sekiguchi, M. Mitome, and D. Golberg, Advanced Materials **19**, 2141 (2007).

[24]  M. Li, L. Xu, C. Sun, Z. Ju, and Y. Qian, Journal of Materials Chemistry **19**, 8086 (2009).

[25]  R. Arenal, M. Kociak, A. Loiseau, and D. J. Miller, Applied Physics Letters **89**, 073104 (2006).

**Figures and Captions**

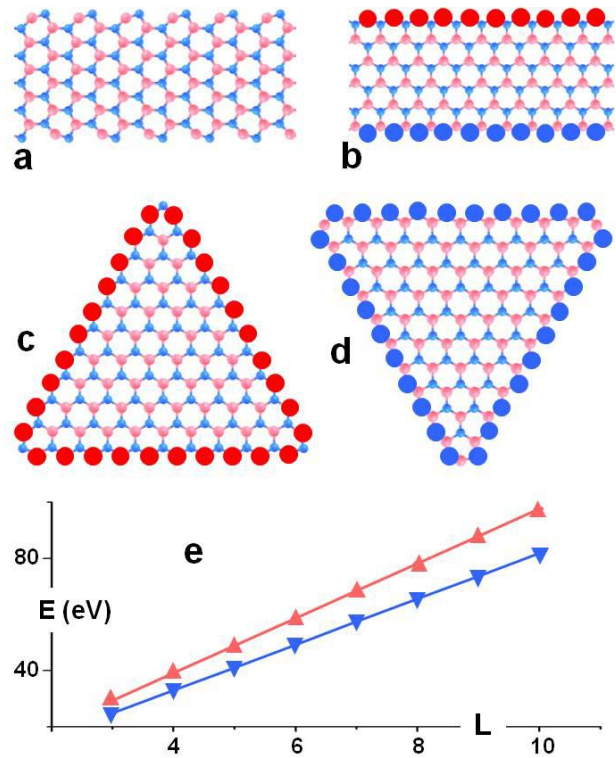

**Figure 1.**

Relaxed geometries of selected BN structures, with boron atoms red and nitrogen blue: (a) an armchair ribbon, note the edge N-atoms buckle out; (b) zigzag ribbon, highlighted B and N at the opposite edges; (c) B-rich triangle and (d) N-rich triangle, of size L = 10. (e) Total energy in a series of triangles shown as a function of their size L, red for B-rich and blue for N-rich.

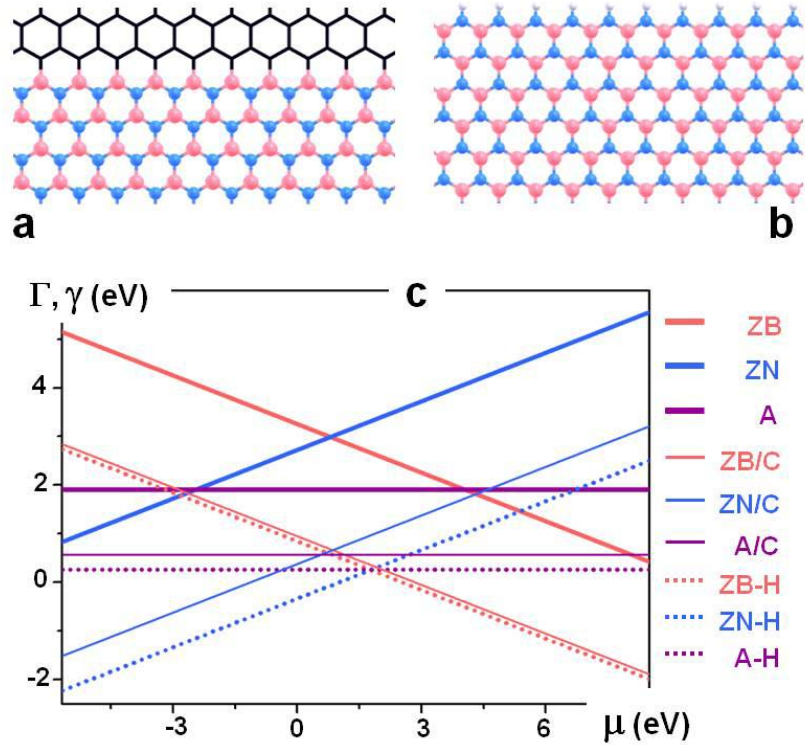

**Figure 2.**

(a) ZB/C interface, with graphene bonds in black. (b) ZN-H edge passivated by hydrogen, white atoms. (c) The interface energy as a function of chemical potential of B: thick lines for bare edges, red is ZB, blue is ZN, purple is A edge; thin lines are for interfaces, red for ZB/C, blue for ZN/C, purple for A/C; dotted lines are red for ZB-H, blue for ZN-H, and purple for A-H.

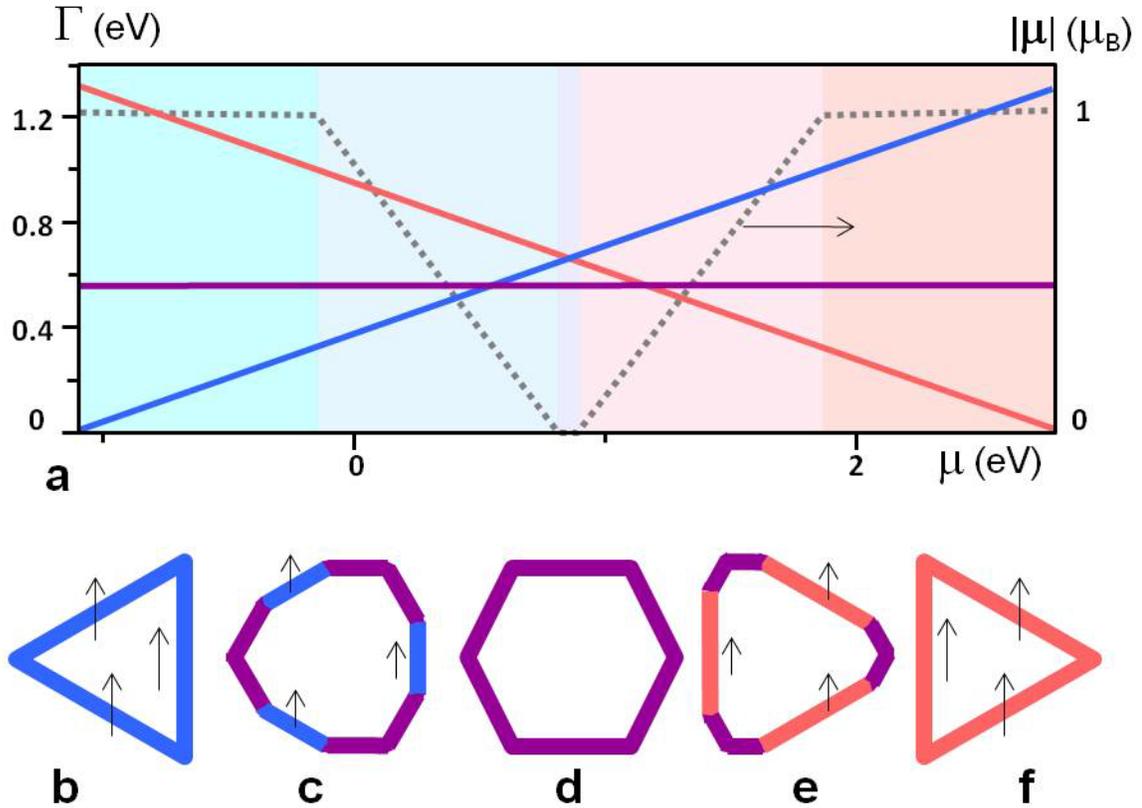

**Figure 3.**

(a) The energies for the boron-rich zigzag (ZB/C, red), armchair (A/C, purple) and nitrogen-rich zigzag (ZN/C, blue) interfaces, as a function of chemical potential of B. Dotted line shows the magnetism, per unit of perimeter, in Bohr magnetons $\mu_B$, as it changes along with the equilibrium shape of graphene inclusions shown in (b-f), from triangle at N-rich condition, to a nonagon (c) and nonmagnetic hexagon (d), and further to inverted nonagon (e) and B-rich triangle again (f). The outline colors mark the interface composition, red for ZB/C, purple for A/C, and blue for ZN/C, and the shapes are computed for $\mu$ = -0.86, 0.42, 0.85, 1.55, and 2.69 eV. Thin arrows length is proportional to the magnitude of magnetism.

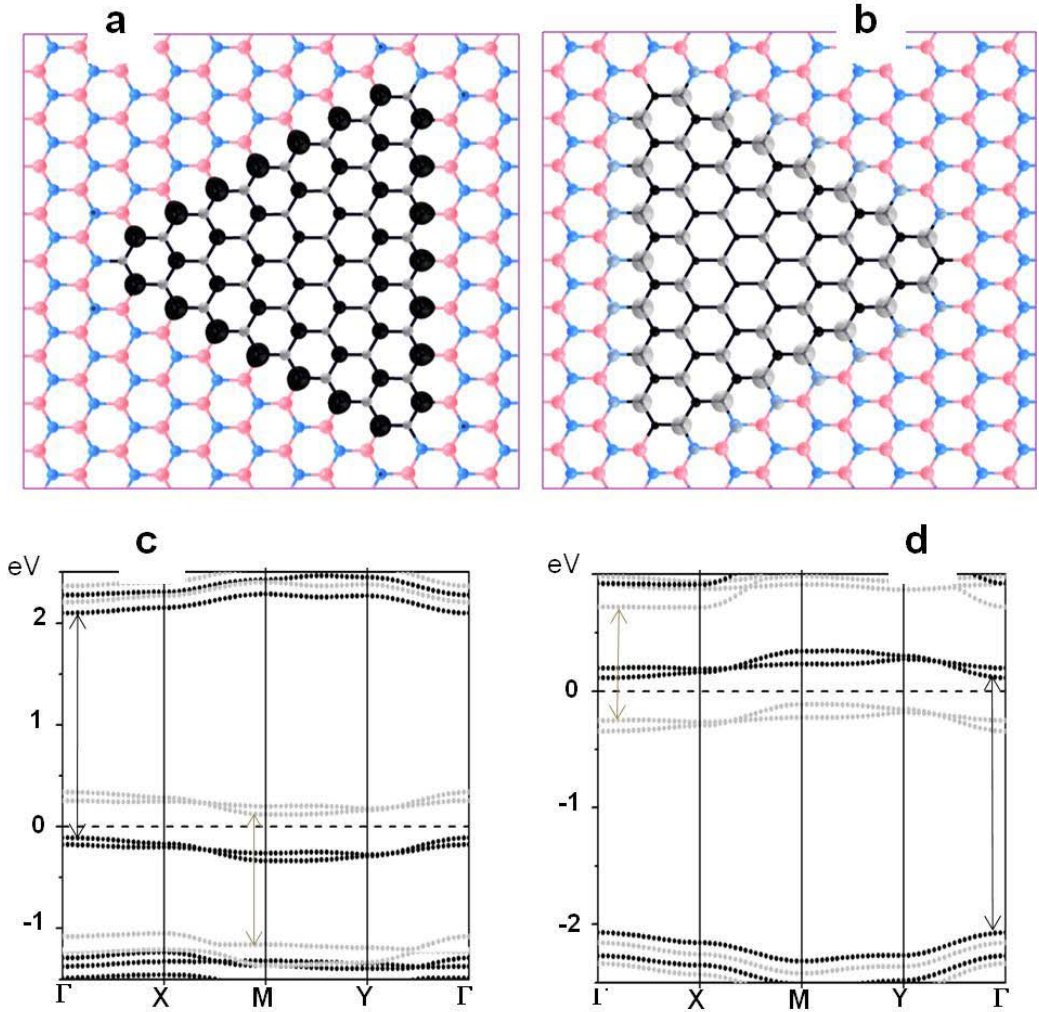

**Figure 4.**

Spin density (a-b) and band structure (c-d) of graphene triangle embedded in BN matrix. Data in (a) and (c) corresponds to boron-rich borders B/C, while (b) and (d) corresponds to nitrogen-rich border N/C. In (a) and (b), dark isosurface is for the density of spin up and gray for the spin down, with isolevel set at 1/10 of maximum. The energy bands in (c) and (d) are plotted along high symmetry directions of rectangular Brillouin zone for graphene inclusion QD of size L = 3, smaller than in (a-b), embedded in a rectangular BN unit cell. G, X, M, Y indicate (0, 0), (1, 0), (1, 1) and (0, 1) points. Black and gray dotted curves show the bands for the opposite spins. Dashed lines show the Fermi level, and the vertical arrows show the band gaps.